# Spectral Characteristics of Large-Scale Radio Emission Areas in Coronal Holes

D. V. Prosovetsky[a], I. Yu. Grigor'eva[b], and A. A. Kochanov[a]

[a] *Institute of Solar—Terrestrial Physics, Siberian Branch, Russian Academy of Sciences, P.O. Box 291, Irkutsk, 664033 Russia*
[b] *Main Astronomical (Pulkovo) Observatory, Russian Academy of Sciences, Pulkovskoe sh. 65, St. Petersburg, 196140 Russia*
e-mail: proso@iszf.irk.ru

**Abstract**—The spectra of the coronal hole radio emission in solar cycles 23 and 24 have been studied based on RATAN-600 data in the 4—16.5 GHz range at frequencies of 5.7 and 17 GHz and 327 MHz. It has been found that bright features of coronal hole microwave emission at 17 GHz and dark features at 5.7 GHz can exist in coronal holes when the spectral index is 1.25—1.5 in the 6.5—16.5 GHz range; the radio spectrum in this range is flat when coronal holes are indiscernible against the background of a quiet Sun. The possible vertical scale of the solar atmosphere over coronal holes is discussed.

## 1. INTRODUCTION

A comparison of coronal hole (CH) radio observations, performed using different instruments at frequencies varying from 150 MHz to 100 GHz, indicate that bright and dark formations can be observed or can be indiscernible against the background of a quiet Sun at different frequencies (see, e.g., the reviews in (Moran et al., 2001; Maksimov et al., 2006)). Increased radioactivity of CH at high frequencies may be explained by wave energy release at the level of upper chromospheres and transition areas (Aihua et al., 1989; Maksimov et al., 2002). However, the CH radio spectral profile is still known insufficiently. The SSRT, Nobeyama, and Nançay radioheliographs are constantly used to determine the radio emission distributions over the solar disk with a high spatial resolution, which makes it possible to identify CHs. An instrument with a wide spectral range—RATAN-600—makes it possible to replenish the deficiency of observational frequencies, as a result of which it becomes possible to determine the frequencies with an increased or decreased emission and, consequently, the characteristic heights where the atmosphere above CHs is additionally heated.

RATAN-600 was previously used to study CHs at frequencies of 1—15 GHz; however, an increased level of CH radio brightness at high frequencies was not found (Borovik et al., 1990). A similar result was achieved in (Papagiannis and Baker, 1982) based on the episodic data of several instruments. At the same time, long-term Nobeyama observations made it possible to find that CHs at 17 GHz are bright formations (Gopalswamy et al., 2000). CH observations at 17 GHz indicated that the averaged brightness temperature of the solar polar regions, where CHs constantly exist, was lower for the period of the cycle 23/24 minimum as compared to the cycle 22/23 minimum (Gopalswamy et al., 2000). The disagreement between the results achieved in (Moran et al., 2001; Maksimov et al., 2006) and (Borovik et al., 1990; Papagiannis and Baker, 1982) can probably be caused by differences in the atmospheric structure above different CHs and changes in the atmospheric properties above CHs during a cycle or between different solar cycles. The CH contrast relative to a quiet Sun at different frequencies can be explained by the relationship between the $T$ and $N_e$ values, which are responsible for the CH thermal emission (Krissinel et al., 2000). Therefore, changes in the CH radio emission properties can probably be caused by a different vertical distribution of these plasma parameters.

## 2. OBSERVATIONS AND THEIR PROCESSING

To study the CH spectral properties, we selected the periods April—June 2012, August 1999, and July 2000. To study the specific features of the CH radio emission, we used two-dimensional (2D) radio brightness distributions obtained at 5.7 (Siberian solar radio telescope, SSRT), 17 GHz (Nobeyama radioheliograph, NoRH), 325, and 150 MHz (Nançay radioheliograph); radio brightness distributions in the 4—16.5 GHz range according to the RATAN-600 data; temperature *(T)* and emission measure (EM) obtained using the *SolarSoft* package, the CHIANTI database, and SOHO/EIT observations. During the selected observation periods, active regions and dark filaments were outside of the RATAN-600 vertical beam, which made it possible to determine the radio flux from large-scale CH areas corresponding to the specific 2D radio brightness distribution in CHs according to the observations at 5.7 and 17 GHz.

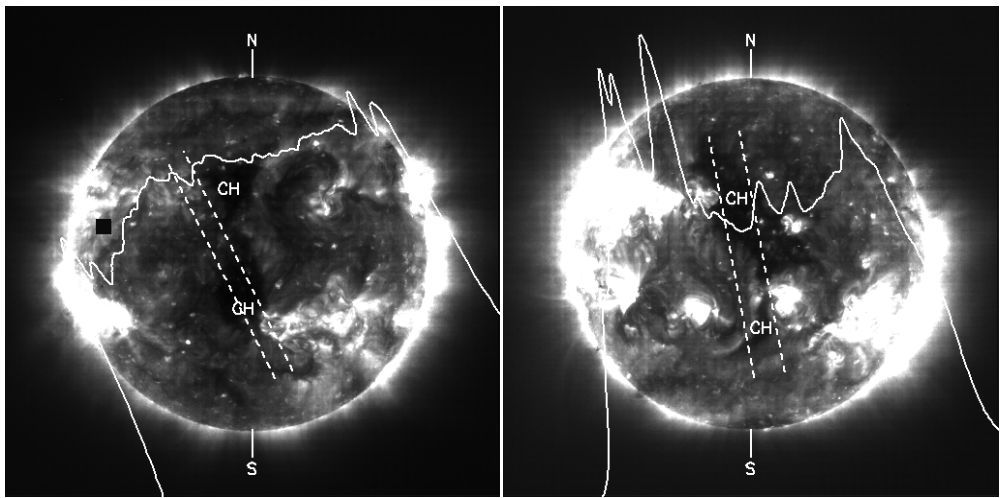

**Fig. 1.** Solid line against the background of an image in the 195 A line shows the radio brightness distribution at a frequency of 5.7 GHz according to the RATAN-600 data. Dotted lines show the CH areas where the radio fluxes were determined. The CH areas on April 10, 2012 (left), and June 3, 2012 (right), are marked with the letters CH.

Figure 1 present's solar images in the 195 Å line according to the EIT data of the SOHO spacecraft with superposed one-dimensional RATAN-600 radio brightness distributions at a frequency of 5.7 GHz and rotated to the position angle. Dotted lines show the solar disk areas where the radio flux at different frequencies was determined for April 10 and June 3, 2012. The areas were distinguished when the radio emission of active regions, flocculi, and dark filaments was minimal according to the data in the 195 Å line and the radio emission from such objects was absent in the regions where the flux was determined in the areas of CH scanning by the RATAN-600 beam on 2D images at 5.7 and 17 GHz (Figs. 2, 3).

## 3. RESULTS OF OBSERVATIONS

The coronal holes to be studied were observed in cycle 23 and during the rising phase of cycle 24. In cycle 23, at a minimum between cycles 23 and 24, and during the rising phase of cycle 24, CHs in the radio emission at 17 and 5.7 GHz manifested themselves differently. During the rise and maximum of cycle 23, CH emission or depression at 17 and 5.7GHz were present (but were nonuniform) within CHs in the UV or X-ray emissions, respectively. At a minimum between cycles 23 and 24 and during the initial rising phase of cycle 24, the CH emission occupied only a small area or was absent within CHs. Coronal holes looked like a radio depression at 327 MHz in cycles 23 and 24 as, well as at a minimum between these cycles and were not detected against the background of a quiet Sun.

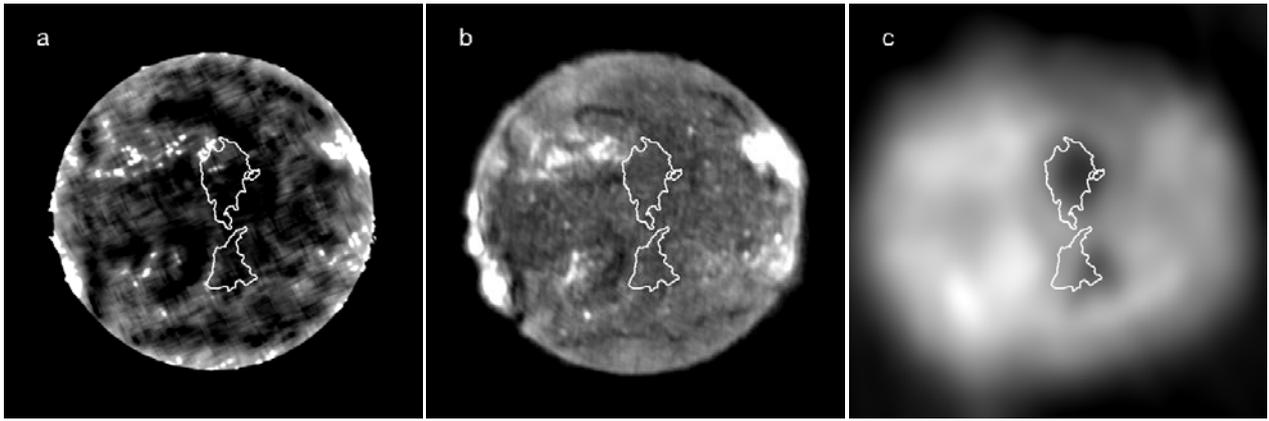

**Fig. 2.** Radio brightness distribution over the solar disk at (a) 17 GHz, (b) 5.7 GHz, and (c) 327 MHz on April 10, 2012. The CH boundaries according to the SOHO/EIT 195 A data are outlined.

During the rising phase of cycle 24, the radio emission at 5.7 and 17 GHz changed in a long-lived mid-latitude CH between April and June 2012. Figure 2 indicates that a CH was not detected against the background of a quiet Sun at frequencies of 17 and 5.7 GHz in April 2012. In June 2012, distinct areas of depression and increased radio brightness, which occupied a substantial part of CHs, appeared at 5.7 and 17 GHz, respectively (Fig. 3). Taking into account the observations at 17 and 5.7 GHz, we can expect that the CH flux from 17 to 5.7 GHz should have values almost equal to the quiet Sun flux for identical areas in April 2012; at the same time, in June 2012, these values should be larger than the quiet Sun flux at 17 GHz, smaller than this flux at 5.7 GHz, and equal to this flux at an intermediate frequency.

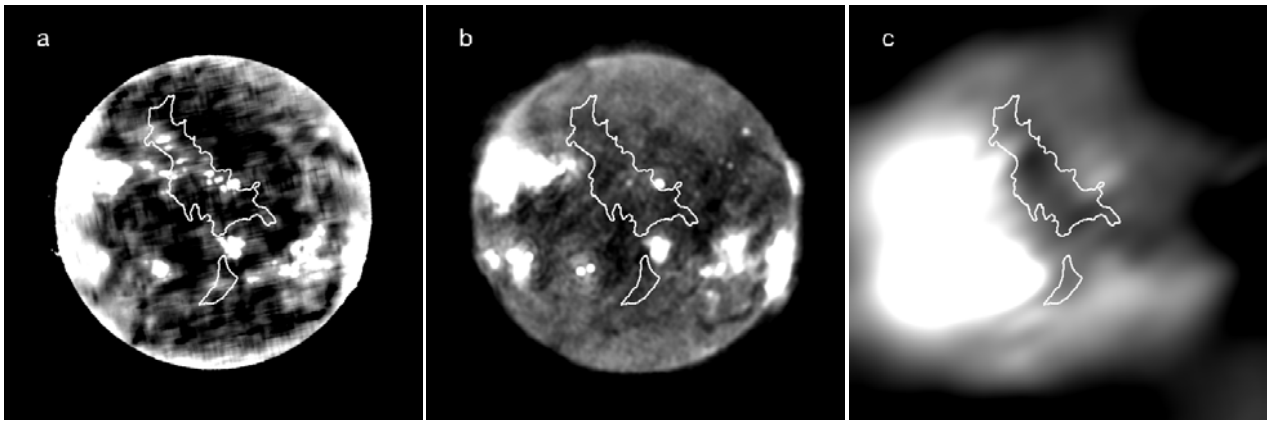

**Fig. 3.** Radio brightness distribution over the solar disk at (a) 17 GHz, (b) 5.7 GHz, and (c) 327 MHz on June 3, 2012. The CH boundaries according to the AIA/SDO 193 A data are outlined.

The CH spectral characteristics in April and June 2012, determined from the RATAN-600 data are shown in Fig. 4. Figure 4 indicates that the CH spectrum in April 2012 is almost flat and slightly differs from the comparable quiet Sun flux; in June 2012 the spectrum shows a pronounced increase at 6.5—16.5 GHz (the spectral index is ~1.6). The CH flux increases in the LF part of the spectrum (<6.5 GHz). The CH flux is equal to the quiet Sun flux at a frequency of ~10 GHz. The RATAN-600 scan section, which was used to obtain the spectrum presented in Fig. 3, corresponds to a large-scale depressed emission at 5.7 GHz and to enhanced emission at 17 GHz in the CH northern area (see Figs. 3a, 3b). Figure 4 presents the values of the radio flux at 5.7 and 17 GHz from the areas for which this flux was determined based on the RATAN-600 data. The values obtained using three instruments are close to one another at the corresponding frequencies. The CH radio fluxes, corresponding to those for which the RATAN-600 flux was determined, were estimated based on the Nançay radioheliograph data at 327 MHz. These fluxes were 0.15 and

0.2 sfu for April 10 and June 3, 2012, respectively. As was mentioned above, CHs at 150 MHz are not detected against the background of a quiet Sun.

The spectra of the CHs that were observed in August 1999 (this spectrum is shown in Fig. 4) and July 2000 are similar to the CH spectrum of June 2012. They have regions with decreased and increased emissions at frequencies higher or lower than 8 GHz and spectral indices of 1.25 and 1.5 in the 6.5—16.5 GHz range.

Based on the 171—195 Å pair of EIT/SOHO lines, which are formed at the boundary between the transition region and corona, it was found that the emission measure (EM) and temperature (T) were $10^{26}$ cm$^{-5}$ and 700 x $10^3$ K in the solar disk areas with depressed and enhanced emission at 5.7 and 17 GHz on June 3, 2012. On April 10, 2012, the EM was 9 x $10^{25}$ cm$^{-5}$ and T was 690 x $10^3$ K in CHs. The differential emission measure (DEM = ΔEM/ΔT) in CHs, determined from two pairs of lines in 171—195 Å and 195—284 Å, was 2 x $10^{20}$ and 7 x $10^{20}$ cm$^{-5}$ K$^{-1}$ for the CHs of June 3 and April 10, 2012, respectively.

## 4. CONCLUSIONS

Based on radio observations of CHs we can conclude that their emission at 150 MHz does not differ from a quiet Sun: the CH radio brightness and, correspondingly, the radio flux slowly decrease with increasing frequency, start increasing between 5 and 7 GHz and over ~9—10 MHz, and exceed the quiet Sun flux in comparable areas; i.e., CH becomes bright in radio. At the same time, CHs that are not visible at 5.7 and 17 GHz are not detected at 150 MHz, are a depression of the radio emission at frequencies of 236—450 MHz (Mercier and Chamble, 2009) and probably near this range, and are not distinguished against the background of a quiet Sun at frequencies varying from 4 to 17 GHz.

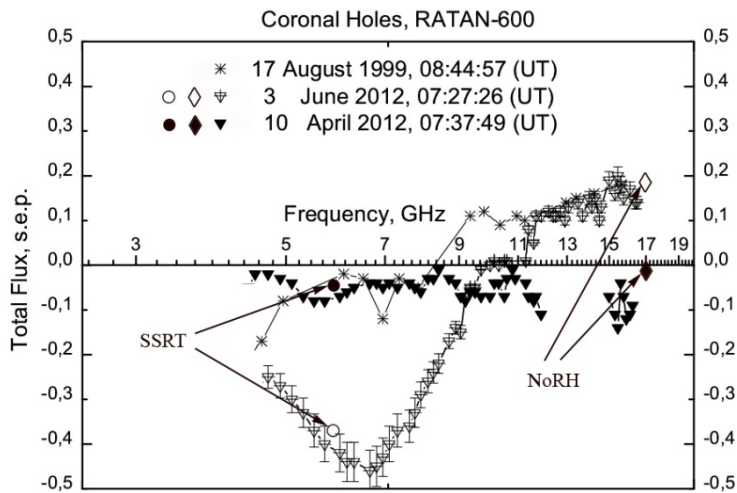

*Fig.* 4. Spectrum of relative radio fluxes in CH areas on August 17, 1999; April 10, 2012; and June 3, 2012. The negative and positive values correspond to CH radio emission depression relative to a quiet Sun and to an increased radio flux, respectively. The CH radio fluxes obtained from the SSRT and NoRH data are marked with circles and diamonds, respectively.

Since the radio emission of CHs is bremsstrahlung, the character of their spectrum in the radio range evidently reflects the vertical plasma temperature and density distribution over CHs. In an optically thin case, the brightness temperature is $T_B \approx T\tau = \frac{9.78 N_e^2 \Lambda}{\nu^2 T^{\frac{1}{2}}} L$, where $N_e$ is the plasma density, $T$ is temperature, ν is the radio frequency, Л is the Coulomb logarithm, and $L$ is the emitting layer thickness. As was shown in Section 3, according to the SOHO/EIT data, the density and temperature values at the boundary between the transition layer and corona in CHs are approximately identical for the cases with a flat spectrum (April 10, 2012) and with a spectrum tilted in the 6.5—16.5 GHz range (June 3, 2012); a difference in the brightness temperature *($T_B$)* between these CHs can be caused by a different thickness of the emitting layer. This is also confirmed by the DEM values, which reflect the vertical scale of variations in the *T* and *N* plasma parameters and differ by a factor of more than 3 in these CHs.

Thus, based on the studied data, we can assume that the CH spectrum depends on the vertical scale of the atmosphere above CHs; the CH emission at frequencies of 17 (brightening) and 5.7 GHz (depression) can exist when the spectral index is ~1.25—1.6 in the 6.5—16.5 GHz range. Taking also into account the fact that the CH emission at 5.7 and 17 GHz occupied the entire CH area in cycle 23 and only partially corresponded to the CH area or was altogether absent between cycles 23 and 24, we can assume that the CH radio emission spectrum depends on the solar cycle

## ACKNOWLEDGMENTS

This work was supported by the RF Ministry of Education and Science (GS 8407 and GK 14.518.11.7047) and by the Russian Foundation for Basic Research (project 12-02-31746 mol_a and 12-02-33110 mol_a_ved).

## REFERENCES

Aihua, Z., Daxiong, F., Jianmin, W., and Chunmei, L., Alfve'n wave heating of the solar atmosphere in the transition region, *Adv. Space Res.,* 1989, vol. 9, issue 4, pp. 33—36.

Borovik, V.N., Kurbanov, M.S., Livshits, M.A., and Ryabov, B.I., Coronal holes against the background of the quiet Sun - observations with the RATAN-600 in the 2—32-cm range, *Sov. Astron.,* 1990, vol. 34, no. 5, p. 522.

Gopalswamy, N., Shibasaki, K., and Salem, M., *J. Astro-phys. Astron.,* 2000, vol. 21, pp. 413—417.

Gopalswamy, N., Yashiro, S., Makela, P., Shibasaki, K., and Hathaway, D., Behavior of solar cycles 23 and 24 revealed by microwave observations, *Astrophy. J. Lett.,* 2012, vol. 750, no. 2, p. L42.

Krissinel, B.B., Kuznetsova, S.M., Maksimov, V.P., Pros-ovetsky, D.V., Grechnev, V.V., Stepanov, A.P., and Shishko, I.F., Some features of manifestations of coronal holes in microwave emission, *Publ. Astron. Soc. Japan,* 2000, vol. 52, p. 909.

Maksimov, V.P. and Prosovetsky, D.V., Coronal heating in the coronal holes regions, *ESA SP-506,* 2002, vol. 2, p. 689.

Maksimov, V.P., Prosovetsky, D.V., Grechnev, V.V., Kris-sinel, B.B., and Shibasaki, K., On the relation of brightness temperatures in coronal holes at 5.7 and 17 GHz, *Publ. Astron. Soc. Japan,* 2006, vol. 58, no. 1, pp. 1—10.

Mercier, C. and Chambe, G., High dynamic range images of the solar corona between 150 and 450 MHz, *Astrophys. J.,* 2009, vol. 700, p. L137.

Moran, T., Gopalswamy, N., Dammasch, I.E., and Wil-helm, K., A multi-wavelength study of solar coronal-hole regions showing radio enhancements, *Publ. Astron. Soc. Japan,* 2001, vol. 378, p. 1037.

Papagiannis, M.D. and Baker, K.B., Determination and analysis of coronal hole radio spectra, *Sol. Phys.,* 1982, vol. 79, p. 365.